\documentclass[american]{article}
\usepackage{mathptmx}
\usepackage[T1]{fontenc}
\usepackage[latin9]{inputenc}
\usepackage{graphicx}

\makeatletter

\providecommand{\tabularnewline}{\\}

\makeatother

\usepackage{babel}

\begin{document}

\title{Hierarchical Complexity: Measures of High Level Modularity}

\author{Alex Fernández (alejandrofer@gmail.com)}
\maketitle
\begin{abstract}
Software is among the most complex endeavors of the human mind; large
scale systems can have tens of millions of lines of source code. However,
seldom is complexity measured above the lowest level of code, and
sometimes source code files or low level modules. In this paper a
hierarchical approach is explored in order to find a set of metrics
that can measure higher levels of organization. These metrics are
then used on a few popular free software packages (totaling more than
25 million lines of code) to check their efficiency and coherency.
\end{abstract}

\section{Introduction}

There is a high volume of literature devoted to measuring complexity.
Several metrics have been proposed to measure complexity in software
packages. At the code level there is for example McCabe's cyclomatic
complexity \cite{key-18}. Complexity in modules (sometimes equated
to source code files, in other occasions low level aggregations of
code) has also been studied, and its relationship to quality has been
measured \cite{key-19}.

But modern software packages can have millions or tens of millions
of source code; just one level of modularity is not enough to make
complex or large software systems manageable. And indeed an exploration
of large system development (made possible thanks to open development
of free software) systematically reveals a hierarchy of levels that
contains subsystems, libraries, modules and other components. But
to date no theoretical approach has been developed to measure and
understand this hierarchy. This paper presents a novel way to study
to this problem; the particular set of metrics derived can be seen
as a first order approach, to be refined in subsequent research when
the problem reaches a wider audience.

\subsection{Contents}

This first section contains this short introduction; the second is
devoted to the study of hierarchical complexity and modularity from
a theoretical point of view. Section 3 explains the experimental setup
that will be used for measurements. In section 4 the source code of
several software packages is measured to verify the theoretical assumptions,
and the results are analyzed. Finally, the last section contains some
conclusions that can be extracted from this approach, in light of
the theoretical analysis and the experimental results.

\subsection{Conventions}

Portions of source code are shown in a non-proportional font: \texttt{org.eclipse}.
File names are also shown in non-proportional type, and also in italics:
\texttt{\emph{org/eclipse}}. Commands are shown indented in their
own line, like this:
\begin{quotation}
\texttt{find . ! -name \textquotedbl{}{*}.c\textquotedbl{} -delete}
\end{quotation}
Terms being defined appear within single quotes: `exponential tree',
while terms already defined in the text or assumed are in double quotes:
{}``combinatorial complexity''.

\section{Theoretical Approach}

Complexity is the bane of software development. Large scale systems
routinely defy the understanding and comprehension powers of the brightest
individuals; small scale systems suffer from poor maintainability,
and it is often cheaper to rewrite old programs than to try to understand
them.

Against this complexity there is basically one weapon: modularity.
No other technique (cataloging, automation, graphical environments)
works for large system development, but our tools to measure it are
scarce. In this section a new metric for complexity measurements is
proposed.

\subsection{Hierarchical Complexity}

In 1991 Capers Jones proposed several measures of complexity for software
\cite{key-13}. One of them was {}``combinatorial complexity'',
dealing with the ways to integrate different components; another was
{}``organizational complexity'', this time dealing with the organization
that produces the software. But no attention was given to the fact
that large-scale software is organized in multiple levels.

Several notions related to high level organization have since been
advanced in the literature \cite{key-14,key-17}. However, when the
time comes to measure complexity in a system it is still the norm
to resort on measuring interfaces\cite{key-12}, code-level complexity
\cite{key-15} or just lines of code per module \cite{key-16}. Given
that the real challenge in software engineering does not currently
lie in algorithmic complexity or cyclomatic complexity, but in the
complexity of high level organization, some metrics to explore complexity
in the component hierarchy should be a useful addition to the software
engineering toolbox. The term {}``hierarchical complexity'' seems
a good fit for the magnitude being measured.

\subsection{System and Components}

A `system' can be defined as a global entity, which is seen from the
outside.

A `component' (often called {}``module'') is an internal part of
the system which is relatively independent, and which hides its inner
workings from the rest of the system. It communicates with the rest
of the system via `interfaces': predefined aggregations of behavior.

We can then redefine a `system' as a collection of components and
the interfaces that connect them. Underlying this definition is the
presumption that a system has well defined components; as we will
see, at some level it holds true for any non-trivial software system.

A system can also be divided in subsystems. A `subsystem' is a high-level
component inside a system. In turn, a component can be divided into
subcomponents; for each of them, the container component is their
{}``system''. In other words, any component can be seen as a whole
system, which interfaces with other systems at the same level.

When the system is viewed in this light, a hierarchy of levels emerges:
the main system is made of subsystems, which can often be decomposed
into components, in turn divided into subcomponents. The number of
levels varies roughly with the size of the system; for small software
projects one or two levels can be enough, while large scale developments
often have more than five levels.

\subsection{Articulation\label{sub:Articulation}}

The way that these levels are organized is a crucial aspect of modularity.
It is not enough to keep a neat division between modules; for robust
and maintainable systems to emerge, connections between levels must
follow a strict hierarchy. The alternative is what is commonly called
{}``spaghetti code''.

There are many different names for components at various levels. Depending
on the domain and the depth we can find subsystems, modules, plugins,
components, directories, libraries, packages and a few more. It is
always a good idea to choose a standard set of levels for a given
system and use them consistently. The selection of names should be
done according to whatever is usual in the system's domain, e.g.:
in operating systems {}``subsystems'' are commonly used as first
level, as in {}``networking subsystem''; while in financial packages
{}``modules'' is the common division (as in {}``supply chain module'').
Finally, ambiguities should be avoided: e.g. in this paper {}``software
package'' is used in this paper in the sense of {}``complete software
systems'', so when adding a component level of {}``packages'' the
difference must be noted.

In the lowest levels it is often the programming language itself that
drives (and sometimes enforces) modularity. In object-oriented languages,
classes are containers of behavior that hide their internal workings
(like attributes or variables), and are in fact a perfect example
of components. Procedural languages usually keep some degree of modularity
within source code files: some of the behavior can be hidden from
other files.

Functions are at the bottom of the hierarchy. A `function' (also called
{}``subroutine'' or {}``method'') is a collection of language
statements with a common interface that are executed in sequence.
A function certainly hides its behavior from the outside, so it can
be seen as the lowest level component, or it can be seen as the highest
level of non-components. The first point of view will be chosen here.

Statements (or lines of code) are the low level elements, but they
can hardly be considered components themselves from the point of view
of the programming language: they do not hide their internal workings,
nor have an interface. There is however one sense in which they can
actually be considered components: as entry points to processor instructions.
A line of code will normally be translated into several machine-code
instructions, which will in turn be run on the processor. The line
of code hides the internal workings of the processor and presents
a common interface in the chosen language; if a function call is performed,
the language statement hides the complexity of converting the parameters
if necessary, storing them and jumping to the desired location, and
finally returning to the calling routine. Since in software development
machine-code instructions are seldom considered except for debugging
or optimization purposes, lines of code remain as {}``borderline
components'' and will be considered as units for the purposes of
this study.

Lines of source code can be unreliable as economic indicators \cite{key-1},
but they are good candidates for the simplest bricks out of which
software is made once comments and blank lines are removed. Source
code statements are more reliable but the added effort is not always
worth the extra precision. In what follows the unit for non-blank,
non-comment source code line will be written like this: {}``$\mathsf{LOC}$'',
with international system prefixes such as {}``$\mathsf{kLOC}$''
for a thousand lines and {}``$\mathsf{MLOC}$'' for a million lines.

To capitulate, a function is made of lines of code, and a class is
made of a number of functions (plus other things). Classes are kept
in source code files, and files are combined into low level components.
Components aggregate into higher level components, until at the last
two steps we find subsystems and the complete system.

\subsection{A Few Numbers}

We will now check these theoretical grounds with some numerical estimations.

The {}``magic number'' $7$ will be used as an approximation of
a manageable quantity; in particular the {}``magic range'' $7\pm2$.
It is common to use the interval $7\pm2$ in that fashion \cite{key-2,key-7},
and yet Miller's original study \cite{key-3} does not warrant it
except as the number of things that can be counted as a glance. Even
so, it seems to be a reasonable cognitive limit for the number of
things that can be considered at the same time.

The main assumption for this subsection is thus that a system is more
maintainable if it is composed of $7\pm2$ components. In this range
its internal organization can be apprehended with ease. Each of these
components will in turn contain $7\pm2$ subcomponents, and so on
for each level. In the lowest level, functions with $7\pm2\mathsf{LOC}$
will be more manageable. In this fashion systems with $7\pm2$ subsystems,
packages with $7\pm2$ classes, classes with $7\pm2$ functions and
functions with $7\pm2\mathsf{LOC}$ will be preferred. At each level
a new type of component will be chosen; arbitrary names are given
so as to resemble a typical software organization.

The problem can thus be formulated as follows: if each component contains
$7\pm2$ subcomponents, what would be the total system size (measured
in $\mathsf{LOC}$) for a given depth of levels? Put another way,
how many levels of components will be needed for a given $\mathsf{LOC}$
count? (Again, the lowest level containing $\mathsf{LOC}$ themselves
does not count as a level of components.)

Let us first study the nominal value of the range, $7$. For only
$3$ levels of components, a reasonable amount for a small system,
the total $\mathsf{LOC}$ count will be \[
7\mathsf{packages}\times7\mathsf{\frac{classes}{package}}\times7\mathsf{\frac{functions}{class}}\times7\mathsf{\frac{LOC}{function}}=2401\mathsf{LOC}.\]

When there are $4$ levels of components, including in this case {}``subsystems'',
the result is \[
7\mathsf{subsystems}\times7\mathsf{\frac{\mathsf{packages}}{subsystem}}\times7\mathsf{\frac{classes}{package}}\times7\mathsf{\frac{functions}{class}}\times7\mathsf{\frac{LOC}{function}}=16807\mathsf{LOC}.\]

For bigger systems a new level of {}``modules'' is added: \[
7\mathsf{subsystems}\times7\mathsf{\frac{\mathsf{modules}}{subsystem}}\times7\mathsf{\frac{\mathsf{packages}}{module}}\times7\mathsf{\frac{classes}{package}}\times7\mathsf{\frac{functions}{class}}\times7\mathsf{\frac{LOC}{function}}=\]

\[
=117649\mathsf{LOC}\approx117\mathsf{kLOC}.\]
A respectable system of $100\mathsf{kLOC}$ is already reached with
$5$ levels. At depth $6$ (inserting a level of {}``libraries'')
the total $\mathsf{LOC}$ count will now be: \[
7\mathsf{subsystems}\times7\mathsf{\frac{\mathsf{libraries}}{subsystem}}\times7\mathsf{\frac{\mathsf{modules}}{library}}\times7\mathsf{\frac{\mathsf{packages}}{module}}\times\ldots\times7\mathsf{\frac{LOC}{function}}=\]

\[
=7^{7}\mathsf{LOC}=823543\mathsf{LOC}\approx824\mathsf{kLOC}.\]

The limit of reason ability can again be estimated using the {}``magic
number'' as $7$ levels, adding a last level of {}``subprojects'':
\[
7\mathsf{subsystems}\times7\mathsf{\frac{\mathsf{subproject}}{subsystem}}\times7\mathsf{\frac{\mathsf{libraries}}{subproject}}\times7\mathsf{\frac{\mathsf{modules}}{library}}\times\ldots\times7\mathsf{\frac{LOC}{function}}=\]

\[
=7^{8}\mathsf{LOC}=5764801\mathsf{LOC}\approx5765\mathsf{kLOC}\approx5.8\mathsf{MLOC}.\]

There are of course systems bigger than 6 million lines of code. A
few ways to extend $\mathsf{LOC}$ count can be noted:
\begin{itemize}
\item Some particular level might contain quite more than $7$ subcomponents.
For example, when classes are counted some of them (those in auxiliary
or {}``scaffolding'' code) can be left out, since they do not belong
to the core of the model and do not necessarily add complexity.
\item We have seen that $\mathsf{LOC}$s are not components, so the $7\pm2$
rule of thumb would not apply at the lowest level. Having e.g. 21
$\mathsf{\frac{LOC}{function}}$ would yield a total three times higher
for every depth.
\item Using the high value in the {}``magic range'' $7\pm2$, $9$, total
size goes up considerably. At the highest depth considered above,
with $7$ levels, size would be $9^{8}=43046721$, or about $43$
million $\mathsf{LOC}$.
\item The {}``magic range'' can also apply to the number of levels, yielding
a total of $9$ levels of components. In this case, $7^{10}=282475249$,
yielding about $282$ million $\mathsf{LOC}$. In the most extreme
high range we would have $9^{10}=3\cdot10^{9}$ or $3486$ million
$\mathsf{MLOC}$, as the most complex system that could be built and
maintained.
\item Lastly, it is quite likely that systems bigger than five million $\mathsf{LOC}$
are indeed too complex to grasp by any single person (or even a team).
A system with five million $\mathsf{LOC}$ is already a challenge
for anyone.
\end{itemize}
It must be noted that the {}``magic range'' does not apply to depth
as such for two different reasons. First, a limit should be sought
rather than a range. Small systems cannot be expected to fall within
the range, if they don't have enough components as to warrant a minimum
of 5 levels. Therefore values below the range are acceptable too;
it is the upper end of the range where it becomes relevant.

A more important objection is that the {}``magic range'' was introduced
above as a cognitive limit for things which have to be considered
at the same time. But component levels are introduced precisely to
avoid considering too many components at the same time; much less
levels of components. There is no real reasoning where all levels
of components have to be considered at the same time, other than to
catalog them; therefore, systems with more than $7\pm2$ levels of
components do not have to be intrinsically harder to manage than systems
with less, other than they become really big at that depth.

\subsection{Generalization\label{sub:Generalization}}

Let $d$ be the depth of the component hierarchy (i.e. the number
of levels of components), and $c(i)$ the number of elements at level
$i$; for $i>0$, $c(i)$ is also the number of subcomponents. Then
the total size $S$ (in $\mathsf{LOC}$) will be \begin{equation}
S=\prod_{i=0}^{d}c(i),\label{eq:total-size}\end{equation}
where $c(0)$ is the number of $\mathsf{LOC}$ per function and $c(1)$
the number of functions per file. Assuming that $c(i)$ is always
in the magic range: \[
c(i)\approx7\pm2,\]
 from where \[
S\approx(7\pm2)^{d+1}.\]

Table \ref{tab:Number-of-LOC} summarizes the estimated number of
$\mathsf{LOC}$ that result from three values in the range $7\pm2$:
both endpoints and the mean value. They can be considered to be representative
of trivial ($5$), nominal ($7$), and complex systems ($9$). Several
depths of levels are shown.

\begin{table}[th]
\begin{centering}
\begin{tabular}{crrr}
\hline 
depth & trivial ($5$) & nominal ($7$) & complex ($9$)\tabularnewline
\hline
$3$ & $0.6\mathsf{kLOC}$ & $2.4\mathsf{kLOC}$ & $6.6\mathsf{kLOC}$\tabularnewline
$4$ & $3.1\mathsf{kLOC}$ & $16.8\mathsf{kLOC}$ & $59\mathsf{kLOC}$\tabularnewline
$5$ & $15.6\mathsf{kLOC}$ & $118\mathsf{kLOC}$ & $531\mathsf{kLOC}$\tabularnewline
$6$ & $78\mathsf{kLOC}$ & $824\mathsf{kLOC}$ & $4783\mathsf{kLOC}$\tabularnewline
$7$ & $391\mathsf{kLOC}$ & $5765\mathsf{kLOC}$ & $43\mathsf{MLOC}$\tabularnewline
$8$ & $1953\mathsf{kLOC}$ & $40\mathsf{MLOC}$ & $387\mathsf{MLOC}$\tabularnewline
$9$ & $9766\mathsf{kLOC}$ & $282\mathsf{MLOC}$ & $3486\mathsf{MLOC}$\tabularnewline
\hline
\end{tabular}
\par\end{centering}

\caption{\label{tab:Number-of-LOC}Number of LOC that corresponds to several
depths of levels. For each depth the low, mean and high values of
the {}``magic range'' $7\pm2$ are shown; representative of trivial,
nominal or complex systems.}

\end{table}

A system whose code is nearer the trivial end will be easy to understand
and master; not only at the lowest code level, but its internal organization.
On the other hand, a complex system will take more effort to understand,
and consequently to maintain and extend. This effort is not only metaphorical:
it translates directly into maintenance costs. An effort to first
simplify such a system before attempting to extend it might be worth
it from an economic point of view.

\subsection{Measurement}

Lines of source code (however imperfect they may be) can be counted;
function definitions can be located using regular expressions. But
it is not easy to get an idea of the number of components in a system.
The basic assumption for the whole study is that source code will
be structured on disk according to its logical organization; \emph{each
directory will represent a logical component}.

An exact measure of modularity would require a thorough study of the
design documents for the software package: first finding out the number
of component levels, then gathering the names of components at each
level and finally counting the number of components. Unfortunately,
such a study cannot be performed automatically; it would only be feasible
on small packages or after a staff-intensive procedure.

A metric based on directories has the big advantage of being easy
to compute and manipulate. It has the disadvantage of imprecision:
to some degree developers may want to arrange files in a manageable
fashion without regard for the logical structure. Developers may also
have an on-disk organization that does not encapsulate internal behavior
and just keeps files well sorted according to some other formal criteria.
Alternatively a software package might be organized logically into
a hierarchy of components, but lack a similar on-disk organization.

Finally, it is of course perfectly possible to rearrange the source
files in new directories to make the code look modular, without actually
changing its internal structure; organization in directories would
allow trivial cheating if used e.g. as an acceptance validation. Since
this measure is not in use today this last concern does not apply,
but it should be taken into account for real-life uses. In software
developed commercially component hierarchies should always be documented,
and automatic checking should only be used as an aid to visual inspections.

\section{Methodology}

The study is performed on several free software packages. Ubuntu Linux
is used for the analysis, but about any modern Unix (complemented
with common GNU packages) should suffice. For each step, the Bash
shell commands are shown.

\subsection{Package Selection and Downloading}

Several free software packages are selected for study. The requirement
to choose only free software is intended to make this study replicable
by any interested party; source code can be made freely available.
The selection of individual packages was made according to the following
criteria:
\begin{itemize}
\item Every package should be relevant: well known in a wide circle of developers
and users.
\item Every package should be successful within its niche; this ensures
that only code which is more or less in a good shape is chosen.
\item Every package should be big enough for a study on complexity; only
software packages bigger than half a million lines of source code
were selected. On the other hand measures should cover about an order
of magnitude. Therefore the selected packages will be in the range
$500\mathsf{kLOC}$ to $5000\mathsf{kLOC}$.
\item At least two packages for each language should be selected.
\item Only packages with several authors should be chosen, so that individual
coding styles are not prevalent.
\end{itemize}
Overall, packages should be written in different languages, so as
to have a varied sample of programming environments.

Once selected, for each package a fresh copy was downloaded in October
2006 to February 2007 from the public distribution. Those packages
that are offered as a single compressed file are downloaded and uncompressed
in their own directory. Some packages have their own download mechanisms.
For the GNOME desktop, GARNOME was used \cite{key-5}: it downloads
the required tools and builds the desktop.

\subsection{Basic Metrics\label{sub:Basic-Metrics}}

The first step is identifying what the target language is, and keeping
just files in that language. It is difficult to find a software project
this size written in only one language; for this study the main language
of each package (in terms of size) was selected, the rest were discarded.

The file name is used to identify source code files. For C we can
choose to keep just those files ending in \texttt{\emph{.c}}, what
is usually called the {}``extension''. C header files (with the
extension \texttt{\emph{.h}}) are ignored, since they are usually
stored along with the equivalent C source code files. Table \ref{tab:Regular-expressions}
shows the extensions for all languages used in the present study.
Then the rest of the files can be deleted:
\begin{quotation}
\texttt{find . ! -name \textquotedbl{}{*}.c\textquotedbl{} -delete}
\end{quotation}
This command removes all files not in the target language; it also
removes all directories that do not contain any target language files.
This step is repeated as needed in case deletion was not strictly
sequential (i.e. a directory was processed before the items it contains),
until no more directories are removed. Then directories are counted:
\begin{quotation}
\texttt{find . -type d | wc -l}
\end{quotation}
All files in the target language are counted too:
\begin{quotation}
\texttt{find . -name \textquotedbl{}{*}.c\textquotedbl{} | wc -l}
\end{quotation}
Lines of code are counted removing blanks and comments. C++ and Java
are derivatives of C, and therefore share a common structure for comments.
Note that lines consisting of opening and closing braces are excluded
too, in order to remove a certain degree of {}``programming style''
from them.
\begin{quotation}
\texttt{find . -name \textquotedbl{}{*}.c\textquotedbl{} -print0 |
xargs -0 cat}

\texttt{~~| grep -v \textquotedbl{}\textasciicircum{}{[}{[}:space:{]}{]}{*}\$\textquotedbl{}
| grep -v \textquotedbl{}\textasciicircum{}{[}{[}:space:{]}{]}{*}\textbackslash{}{*}\textquotedbl{}}

\texttt{~~| grep -v \textquotedbl{}\textasciicircum{}{[}{[}:space:{]}{]}{*}\{{[}{[}:space:{]}{]}{*}\$\textquotedbl{}}

\texttt{~~| grep -v \textquotedbl{}\textasciicircum{}{[}{[}:space:{]}{]}{*}\}{[}{[}:space:{]}{]}{*}\$\textquotedbl{} }

\texttt{~~| grep -v \textquotedbl{}\textasciicircum{}{[}{[}:space:{]}{]}{*}//\textquotedbl{}
| wc -l}
\end{quotation}
The results obtained with this method are not perfect, but they have
been found to be a reasonable fit. Similar expressions are used for
Lisp and Perl taking into account their respective syntax for comments.
For the actual results presented in this study several equivalent
regular expressions are used to remove comments, with slightly more
accurate results.

The number of functions is also counted using full-blown regular expressions.
Table \ref{tab:Regular-expressions} shows the regular expressions
used for each language to locate function definitions in the code.

\begin{table}[th]
\begin{centering}
\begin{tabular}{llr}
\hline 
language & extension & function definition\tabularnewline
\hline
C & {}``\texttt{.c}'' & \texttt{(\textbackslash{}w+)\textbackslash{}s{*}\textbackslash{}({[}\textbackslash{}w\textbackslash{}s\textbackslash{},\textbackslash{}{[}\textbackslash{}{]}\textbackslash{}\&\textbackslash{}{*}{]}{*}\textbackslash{})\textbackslash{}s{*}\textbackslash{};}\tabularnewline
C++ & {}``\texttt{.cpp}'',{}``\texttt{.cxx}'',{}``\texttt{.cc}'' & \texttt{(\textbackslash{}w+)\textbackslash{}s{*}\textbackslash{}({[}\textbackslash{}w\textbackslash{}s\textbackslash{},\textbackslash{}{[}\textbackslash{}{]}\textbackslash{}\&\textbackslash{}{*}{]}{*}\textbackslash{})\textbackslash{}s{*}\textbackslash{}\{}\tabularnewline
Java & {}``\texttt{.java}'' & \texttt{(\textbackslash{}w+)\textbackslash{}s{*}\textbackslash{}({[}\textbackslash{}w\textbackslash{}.\textbackslash{}s\textbackslash{},\textbackslash{}{[}\textbackslash{}{]}{]}{*}\textbackslash{})\textbackslash{}s{*}\ldots{}}\tabularnewline
 &  & \texttt{(?:throws\textbackslash{}s{*}\textbackslash{}w+(?:\textbackslash{}s{*}\textbackslash{},\textbackslash{}s{*}\textbackslash{}w+){*})\textbackslash{}s{*}\textbackslash{}\{}\tabularnewline
Lisp & {}``\texttt{.el}'' & \texttt{\textbackslash{},\textbackslash{}s{*}\textbackslash{}w+){*}){*}\textbackslash{}s{*}\textbackslash{}\{}\tabularnewline
Perl & {}``\texttt{.pl}'',{}``\texttt{.plx}'',{}``\texttt{.pm}'' & \texttt{(\textbackslash{}w+)\textbackslash{}s{*}\textbackslash{}(;
(?<=sub\textbackslash{}s)\textbackslash{}s{*}(\textbackslash{}w+)\textbackslash{}s{*}\textbackslash{}\{}\tabularnewline
\hline
\end{tabular}
\par\end{centering}

\caption{\label{tab:Regular-expressions}File extensions for different languages.
The right-most column shows the regular expression used to find function
definitions for each language.}

\end{table}

A Python script iterates over every directory and file with the correct
extension, and counts the number of matches for the appropriate regular
expression. Again, the results have been inspected for correctness.
They are not perfect; for example, some C++ function definitions may
call parent functions, and when they contain typecasts they will not
be registered under this definition. C++ template code is not considered
either. They are however a good enough compromise; not too slow but
not too precise.

Note that C and C++ headers are not used to find function definitions.
There are several reasons for this decision. First, header files usually
contain only function definitions, while function declarations are
in source files (except for inlined functions). And second and most
important, it is hard to distinguish between C and C++ headers since
they follow the same pattern (both share the extension {}``\texttt{.h}''),
so they would get mixed in the process for projects that contain code
in both languages. In field experiments the results are not affected
by this omission.

\subsection{Average Directory Depth\label{sub:Average-directory-depth}}

A first approximation to complexity is to find out the depth of each
source code file, and compute an average. This operation can be performed
using a sequence of commands like this (for C code):
\begin{quotation}
\texttt{tree | grep {}``\textbackslash{}.c'' | grep {}``.-{}-''
| wc -l}

\texttt{tree | grep {}``\textbackslash{}.c'' | grep {}``.~~~.-{}-''
| wc -l}

\texttt{tree | grep {}``\textbackslash{}.c'' | grep {}``.~~~.~~~.-{}-''
| wc -l}

\texttt{\ldots{}}
\end{quotation}
thus getting the source file counts for level 0, 1, 2\ldots{} and
computing the differences to obtain the number of files at each level
of directories. An average depth can then be easily computed, where
the depth for each file is the number of directories that contain
it. Table \ref{tab:Depth-of-source} shows the number of source files
at each level for a few packages.

\begin{table}[th]
\begin{centering}
\begin{tabular}{lrrrrrrrrrr}
\hline 
package & 1 & 2 & 3 & 4 & 5 & 6 & 7 & 8 & 9 & 10\tabularnewline
\hline
Linux & 297 & 2474 & 4451 & 1050 & 142 & 0 & 0 & 0 & 0 & 0\tabularnewline
OpenSolaris & 0 & 2 & 1529 & 3686 & 5082 & 1850 & 340 & 104 & 10 & 3\tabularnewline
Eclipse & 2 & 0 & 89 & 228 & 2045 & 4136 & 4674 & 2973 & 407 & 0\tabularnewline
\hline
\end{tabular}
\par\end{centering}

\caption{\label{tab:Depth-of-source}Depth of source code files for selected
software packages.}

\end{table}

\begin{figure}[th]
\begin{centering}
\includegraphics{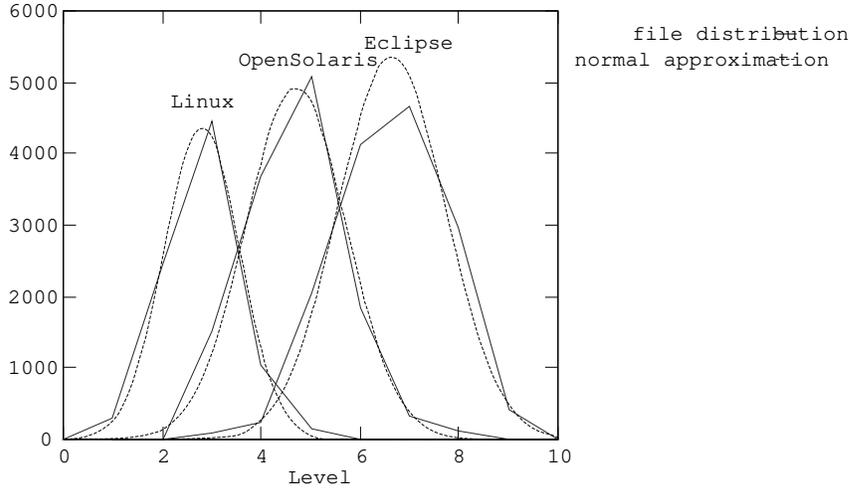}
\par\end{centering}

\caption{\label{fig:Depth-of-source}Depth of source code files organized by
levels. A normal approximation is shown for each package. }

\end{figure}

Figure \ref{fig:Depth-of-source} shows the previous distribution
graphically. Each set of depths can be approximated by a normal distribution.

This approach has several problems. Many packages contain directories
with only one subdirectory, e.g. a directory called \texttt{\emph{source}}
or \texttt{\emph{src}} which contains all source code; this is an
artifact of the source distribution and should not be taken into account.
In other occasions folders are repeated across the board, as in Java
packages: the convention here is to start all packages with a prefix
that depends on the organization that develops the code, and continue
with the name of the organization and possibly some other fixed directories
\cite{key-26}. E.g. for Eclipse all packages should start with \texttt{org.eclipse},
which results in folders \texttt{\emph{org/eclipse}}; this structure
is repeated inside each individual project within the Eclipse source
code. Visual inspection of the result is therefore necessary to disregard
those meaningless directories.

The results are less than satisfactory. In what follows an alternative
approach will be explored.

\subsection{Average Number of Items per Directory}

The target of complexity measurements is to find the number of subcomponents
per component. From the number of archives and the number of directories
a rough approximation to modularity can be computed, as the ratio
between files and directories. But directories can also be contained
in other directories, counting as subcomponents.

Let $T$ be the total number of source files, and $D$ be the number
of directories. Then the average number of items per directory $a$
can be computed as \begin{equation}
a=\frac{T+D}{D+1},\label{eq:at+d}\end{equation}
counting one more directory in the denominator for the root directory.
It is a measure of the average number of subcomponents per component.

\subsection{Average Exponential Depth\label{sub:Average-Exponential-depth}}

Once the average number of items per directory is known, another approach
to compute the average depth can be used.

\begin{figure}[th]
\begin{centering}
\includegraphics{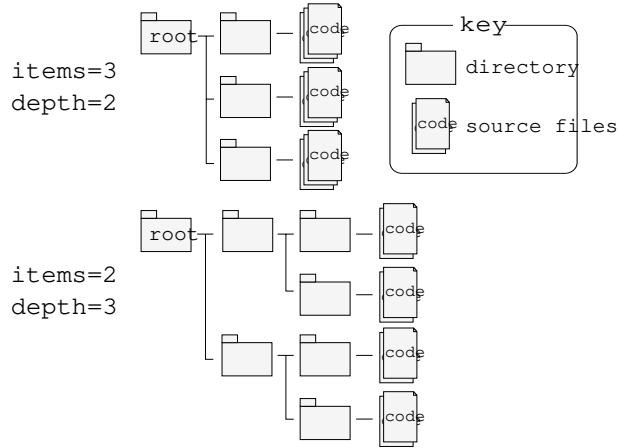}
\par\end{centering}

\caption{\label{fig:Exponential-tree}Two examples of exponential trees. The
number of items (directories and files) per directory at each level
is constant throughout the tree. }

\end{figure}

An `exponential tree' is a tree of directories that has the same number
of items at every level. The last level contains only files; all the
rest contain only directories. Figure \ref{fig:Exponential-tree}
shows two examples of exponential trees: one with 3 items per directory
and 2 levels, for a total of $3^{2}=9$ source code files; and one
with 2 items per directory and 3 levels, yielding $2^{3}=8$ source
files. The root directory is not counted.

Symbolically:\begin{equation}
T=a^{l},\label{eq:exponential-basic}\end{equation}
where $T$ is the total number of files, and $a$ the number of items
per directory. The number of files in the tree grows exponentially
with its depth $l$.

If the tree of source code files and directories is approximated by
an exponential tree, then at each level $i$ the complexity would
be constant: $c(i)=a$. If such a tree has $l$ levels, then equation
\ref{eq:exponential-basic} can be written as: \begin{equation}
T=\prod_{i=2}^{l+1}c(i)=a^{l},\label{eq:exponential}\end{equation}
where levels are counted from $2$ to $l+1$ because level $1$ is
for functions inside files. The component hierarchy will actually
have $l+1$ levels. With some arithmetic we get: \[
l=\frac{ln(T)}{ln(a)},\]
which considering equation \ref{eq:at+d} can be restated as \begin{equation}
l=\frac{ln(T)}{ln(T+D)-ln(D+1)}\label{eq:levels}\end{equation}
and which can be computed just knowing the number of source files
and the number of directories.

This value represents the depth that an exponential tree would need
to have in order to produce the observed number of source code files;
the approximation is evidenced by the presence of fractional depths.
If the overall number of levels follows the $7\pm2$ rule, the value
for exponential depth should cluster around $6$, since it does not
take into consideration the lowest function level. Symbolically: \[
l=d-1\approx6\pm2.\]

\subsection{Pruning\label{sub:Pruning}}

Consolidation can play an important role in some source code layouts.
Each subsystem may have its own hierarchy of directories like \texttt{\emph{src}};
they add to the count of directories but do not really add either
complexity or modularity. This effect can be exacerbated in some Java
projects (particularly those organized around Eclipse projects), where
the common hierarchy of packages (like \texttt{\emph{org.eclipse\ldots{}}})
is repeated for every subproject, inside the \texttt{\emph{src}} folder.
Directories should be consolidated removing repeated occurrences prior
to measuring, but this operation requires intimate knowledge of the
structure of the software and will not be done here.

A more practical approximation to consolidation can be done by pruning
the source code tree. A certain amount of pruning has already been
done in subsection \ref{sub:Basic-Metrics}: non-source files and
empty directories have already been pruned, but a more aggressive
approach is required. The number of reported directories is in many
instances not consistent with .

\begin{figure}[th]
\begin{centering}
\includegraphics{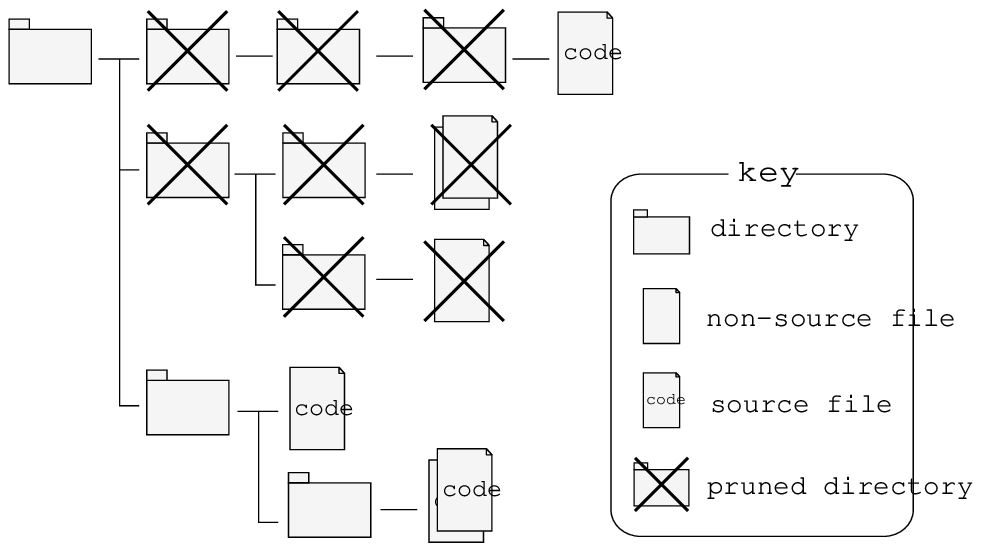}
\par\end{centering}

\caption{\label{fig:Pruning}Pruning of the source code tree. Non-source files,
empty directories and directories containing only one meaningful item
are pruned.}

\end{figure}

`Trivial' directories are those containing just one meaningful item:
either one code file or one meaningful directory. (A `meaningful'
directory is one which contains source code at any level.) These trivial
directories cannot add to the component hierarchy, since a component
with one subcomponent is meaningless; therefore they are not counted
as directories. Figure \ref{fig:Pruning} shows an example of pruning
a source code tree.

In practice, a small Python script (available upon request from the
author) is used to count source files and directories, discount trivial
directories, recalculate the values for $T$ and $D$, and recompute
the average number of items per directory.

\begin{figure}[th]
\begin{centering}
\includegraphics{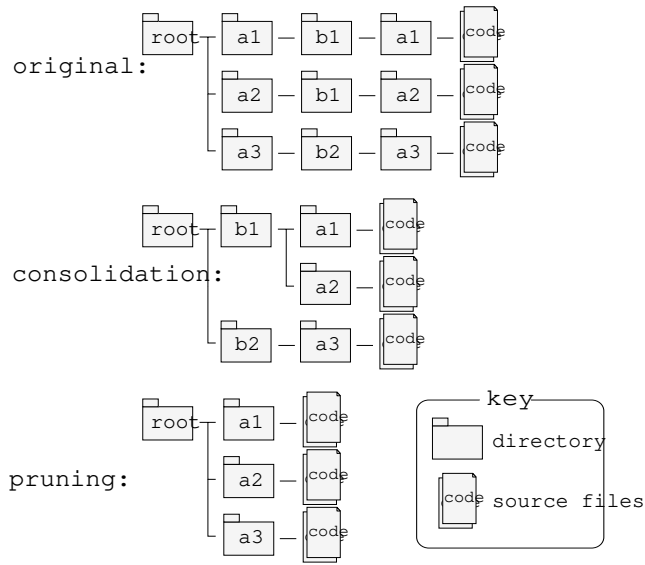}
\par\end{centering}

\caption{\label{fig:Pruning-Consolidating}Example of the results of pruning
compared to consolidation. The original structure shows that directories
$a1-a3$ had an internal linear structure with repeated directories,
yielding 10 directories. Consolidation leads to removing the leading
directory and sharing the common directory $b1$ for $a1$ and $a2$,
for a total of 6 directories. Pruning removes every trivial directory,
leaving only 4 directories.}

\end{figure}

Again, the results are not perfect since pruning does not yield the
same results as consolidation: some internal structure can be lost
in the process, so this time it tends to underestimate the number
of directories, albeit in a smaller percentage. Figure \ref{fig:Pruning-Consolidating}
shows an example of pruning compared to consolidation.

Nevertheless, pruned values seem to follow code organization more
closely than in the original model, so they will be used in the final
results.

\section{Results}

The following tables summarize the results for a selected range of
software projects.

\subsection{Basic Metrics}

Table \ref{tab:Lines-of-code} shows the results of computing some
basic metrics against the selected software packages.

\begin{table}[th]
\begin{centering}
\begin{tabular}{lcrrrr}
\hline 
package & language & $\mathsf{kLOC}$ & functions & files & directories\tabularnewline
\hline
Linux 2.6.18 &  C & 3388 & 134682 &  8414 &  902\tabularnewline
OpenSolaris 20061009 &  C & 4299 & 120925 & 12606 & 2525\tabularnewline
GNOME 2.16.1 &  C & 3955 & 163975 & 10025 & 1965\tabularnewline
KDE 20061023 & C++ & 2233 & 172761 & 11016 & 1602\tabularnewline
KOffice 20061023 & C++ & 511 & 31880 & 2204 & 375\tabularnewline
OpenOffice.org C680\_m7 & C++ & 3446 & 114250 & 11021 & 1513\tabularnewline
SeaMonkey 1.0.6 & C++ & 1180 & 69049 & 4556 & 884\tabularnewline
Eclipse 3.2.1 & Java & 1560 & 163838 & 14645 & 2334\tabularnewline
NetBeans 5.5 & Java & 1615 & 187212 & 16563 & 6605\tabularnewline
JBoss 5.0.0.Beta1 & Java & 471 & 54645 & 9504 & 2478\tabularnewline
Emacs  21.4 & Lisp & 473 & 18931 & 759 & 18\tabularnewline
XEmacs-Packages 21.4.19 & Lisp & 926 & 37415 & 2133 & 285\tabularnewline
\hline
\end{tabular}
\par\end{centering}

\caption{\label{tab:Lines-of-code}Lines of code, function definitions, source
files and directories for selected free software packages.}

\end{table}

It must be noted how non-modular the code of Emacs really is, evident
even in these direct measurements.

\subsection{Average Items per Directory}

These basic metrics allow us to compute some basic complexity metrics.
Table \ref{tab:Basic-complexity-metrics} shows $\mathsf{LOC}$ per
function (an indication of the lowest level of articulation), functions
per file (the first level of components) and items per directory (all
the remaining levels of components).

\begin{table}[th]
\begin{centering}
\begin{tabular}{lrrr}
\hline 
package & $\mathsf{LOC}$/function & functions/file & items/directory\tabularnewline
\hline
Linux 2.6.18 & $25.2$ & $16.0$ & $10.33$\tabularnewline
OpenSolaris 20061009 & $35.6$ & $9.6$ & $5.99$\tabularnewline
GNOME 2.16.1 & $24.1$ & $16.4$ &  $6.10$\tabularnewline
KDE 20061023 & $12.9$ & $15.7$ & $7.88$\tabularnewline
KOffice 20061023 & $16.0$ & $14.5$ & $6.88$\tabularnewline
OpenOffice.org C680\_m7 & $30.2$ & $10.4$ & $8.28$\tabularnewline
SeaMonkey 1.0.6 & $17.1$ & $15.2$ & $6.15$\tabularnewline
Eclipse 3.2.1 & $9.5$ & $11.2$ &  $7.27$\tabularnewline
NetBeans 5.5 & $8.6$ & $11.3$ & $3.51$\tabularnewline
JBoss 5.0.0.Beta1 & $8.6$ & $5.7$ & $4.84$\tabularnewline
Emacs 21.4 & $24.6$ & $25.3$ & $43.17$\tabularnewline
XEmacs-Packages 21.4.19 & $24.7$ & $17.6$ & $8.46$\tabularnewline
\hline
\end{tabular}
\par\end{centering}

\caption{\label{tab:Basic-complexity-metrics}Basic complexity metrics.}

\end{table}

Both $\mathsf{LOC}$ per function and functions per file are shown
separately from the remaining levels, items per directory. And indeed
we find a bigger disparity in $\mathsf{LOC}$ per function than in
the remaining metrics; from the $8.6$ found in JBoss or NetBeans
to the $35.6$ in OpenSolaris. In functions per file both procedural
and functional languages (C and Lisp) appear to be similar to C++;
only Java shows lower values.

In the remaining metric, items per directory, only two packages are
above the {}``magic range'' $7\pm2$: Emacs and Linux. Emacs stands
out as the less modular of all packages with difference, which is
consistent with its long heritage: it is an ancient package which
has seen little in the way of modern practices, and which is not really
modularized. The modernized version, XEmacs, shows a profile more
coherent with modern practices, although only the separate packages
are measured. Linux, on the other hand, is a monolithic kernel design
with a very practical approach. This seems to result in source code
which is not always as well modular as it might be. JBoss is also
out of range by a very small amount, but this time it is the low end
of the range ($4.84$ items per directory). Its code is very modular,
so much so that with only half a million $\mathsf{LOC}$ it has the
second highest directory count. The first position goes to NetBeans,
with $3.51$ items per directory; it has by far the highest directory
count.

These first impressions are refined and expanded in subsection \ref{sub:Discussion}.

\subsection{Average depth}

Average depth is computed using two algorithms: average directory
depth (see subsection \ref{sub:Average-directory-depth}) and average
exponential depth (see equation \ref{eq:levels} in subsection \ref{sub:Average-Exponential-depth}).
Table \ref{tab:Average-depth} shows the average depth for the usual
set of packages, calculated using both methods.

\begin{table}[th]
\begin{centering}
\begin{tabular}{lrr}
\hline 
package & directory & exponential\tabularnewline
\hline
Linux 2.6.18 & $2.8$ & $3.9$\tabularnewline
OpenSolaris 20061009 & $4.7$ & $5.3$\tabularnewline
GNOME 2.16.1 & $3.8$ & $5.1$\tabularnewline
KDE 20061023 & $3.2$ & $4.5$\tabularnewline
KOffice 20061023 & $2.5$ & $3.1$\tabularnewline
OpenOffice.org C680\_m7 & $3.7$ & $4.4$\tabularnewline
SeaMonkey 1.0.6 & $4.8$ & $4.6$\tabularnewline
Eclipse 3.2.1 & $4.6$ & $4.8$\tabularnewline
NetBeans 5.5 & $7.5$ & $7.7$\tabularnewline
JBoss 5.0.0.Beta1 & $4.9$ & $5.8$\tabularnewline
Emacs  21.4 & $1.7$ & $1.8$\tabularnewline
XEmacs-Packages 21.4.19 & $3.0$ & $3.6$\tabularnewline
\hline
\end{tabular}
\par\end{centering}

\caption{\label{tab:Average-depth}Average depth computed using directory depth
and exponential depth.}

\end{table}

The biggest difference between both methods happens in KDE ($3.2$
vs $4.5$), GNOME ($3.8$ vs $5.1$) and Linux ($2.8$ vs $3.9$).
The values of directory depth for the first two packages involve heavy
corrections due to directories with just one subdirectory in the tree;
for Linux however there are no such corrections, but there is a big
disparity between some directories which can contain up to $116$
items, like {}``\texttt{/fs}'', and others with just one item. The
rest of the packages are within a level in both measures. Exponential
depth does not involve error-prone corrections and will therefore
be used in what follows.

This average depth of components does not take into account the lowest
level of components that was considered in subsection \ref{sub:Articulation},
that of functions. To find out the real depth of the component hierarchy
a level must therefore be added to the exponential depth. In fact,
for C++ packages an additional level for classes might also be added,
since they are allowed a further level of modularity inside source
code files. This possibility will be discussed in its own subsection,
\ref{sub:c++}.

\subsection{Pruning}

After pruning the source code tree, disregarding all trivial directories
(those containing only one file or directory), the results are those
in table \ref{tab:Pruned-dirs}.

\begin{table}[th]
\begin{centering}
\begin{tabular}{lrrr}
\hline 
package & directories & items/directory & depth\tabularnewline
\hline
Linux 2.6.18 & 774 & $11.9$ & $3.65$\tabularnewline
OpenSolaris 20061009 & 1276 & $10.9$ & $3.96$\tabularnewline
GNOME 2.16.1 & 1075 & $10.3$ &  $3.95$\tabularnewline
KDE 20061023 & 1444 & $10.2$ & $4.09$\tabularnewline
KOffice 20061023 & 286 & $8.7$ & $3.56$\tabularnewline
OpenOffice.org C680\_m7 & 1134 & $10.7$ & $3.92$\tabularnewline
SeaMonkey 1.0.6 & 532 & $9.6$ & $3.73$\tabularnewline
Eclipse 3.2.1 & 1116 & $14.1$ &  $3.62$\tabularnewline
NetBeans 5.5 & 2492 & $7.6$ & $4.78$\tabularnewline
JBoss 5.0.0.Beta1 & 1471 & $7.5$ & $4.56$\tabularnewline
Emacs 21.4 & 16 & $48.4$ & $1.71$\tabularnewline
XEmacs-Packages 21.4.19 & 120 & $18.8$ & $2.61$\tabularnewline
\hline
\end{tabular}
\par\end{centering}

\caption{\label{tab:Pruned-dirs}After pruning trivial directories: pruned
directory count, average items per directory and exponential depth.}

\end{table}

Comparing these results with those in table \ref{tab:Basic-complexity-metrics}
it can be seen that items per directory have increased (while depth
has correspondingly decreased), as could be expected by the smaller
number of directories considered. Values have mostly gone out of the
{}``magic range'', to the point that only three software packages
remain inside: two in Java, NetBeans and JBoss, and one in C++: KOffice.

\subsection{Complexity per Level}

Table \ref{tab:Corrected-depth} shows the depth once the correction
for functions (adding one level to every depth) is taken into account.
It also summarizes the number of elements per component at levels
0 ($\mathsf{LOC}$ per function), 1 (functions per file) and higher
than 1 (items per directory); according to the assumptions in subsection
\ref{sub:Generalization}, this number would represent complexity
at each level.

\begin{table}[th]
\begin{centering}
\begin{tabular}{lrrrr}
\hline 
package & level 0  & level 1 & level > 1 & depth\tabularnewline
\hline
Linux 2.6.18 & $25.2$ & $16.0$ & $11.9$ & $4.65$\tabularnewline
OpenSolaris 20061009 & $35.6$ & $9.6$ & $10.9$ & $4.96$\tabularnewline
GNOME 2.16.1 & $24.1$ & $16.4$ & $10.3$ &  $4.95$\tabularnewline
KDE 20061023 & $12.9$ & $15.7$ & $10.2$ & $5.09$\tabularnewline
KOffice 20061023 & $16.0$ & $14.5$ & $8.7$ & $3.56$\tabularnewline
OpenOffice.org C680\_m7 & $30.2$ & $10.4$ & $10.7$ & $4.92$\tabularnewline
SeaMonkey 1.0.6 & $17.1$ & $15.2$ & $9.6$ & $4.73$\tabularnewline
Eclipse 3.2.1 & $9.5$ & $11.2$ & $14.1$ &  $4.62$\tabularnewline
NetBeans 5.5 & $8.6$ & $11.3$ & $7.6$ & $5.78$\tabularnewline
JBoss 5.0.0.Beta1 & $8.6$ & $5.7$ & $7.5$ & $5.56$\tabularnewline
Emacs 21.4 & $24.6$ & $25.3$ & $48.4$ & $1.71$\tabularnewline
XEmacs-Packages 21.4.19 & $24.7$ & $17.6$ & $18.8$ & $3.61$\tabularnewline
\hline
\end{tabular}
\par\end{centering}

\caption{\label{tab:Corrected-depth}Complexity at various levels and depth
(number of levels) of the component hierarchy. Complexity is computed
as number of subcomponents per component; corrected depth is found
adding one to exponential depth. Pruning is used.}

\end{table}

$\mathsf{LOC}$ count can be approximately recovered based on these
measures. To recapitulate: $S$ is the size in $\mathsf{LOC}$, and
$c(i)$ is the number of elements at level $i$; $d$ is the total
hierarchy depth (number of levels), $a$ the average number of items
per directory, and $T$ is the total number of source files. Now $l$
is the exponential depth, so that $d=l+1$. (Table \ref{tab:Corrected-depth}
shows, from left to right: $c(0)$, $c(1)$, $a$ and $d$.) Then
starting from equation \ref{eq:total-size}: \[
S=\prod_{i=0}^{d}c(i)=c(0)\times c(1)\times T,\]
and recalling equation \ref{eq:exponential}: \[
S=c(0)\times c(1)\times a^{d-1}.\]
The count is not exact due to rounding.

\subsection{Discussion\label{sub:Discussion}}

In this subsection we will discuss the experimental results, referring
to table \ref{tab:Corrected-depth} unless explicitly stated.

Looking just at corrected depth, almost all measured values are well
below the {}``magic range'' $7\pm2$, as corresponds their medium
size. In fact there are only three values in the range: NetBeans at
$5.78$, JBoss at $5.56$ and KDE with $5.09$, with several others
grazing the $5$. Interestingly, the second deepest hierarchy is found
in JBoss, even though it is the smallest package considered.

Taking now a closer look at complexity for level > 1 (i.e. average
items per directory): the first thing to note is that before pruning
(table \ref{tab:Basic-complexity-metrics}) most values were inside
the {}``magic range''. After pruning there are only three values
inside the range; this time all the rest are above. JBoss holds now
the lowest value with $7.5$, very close to NetBeans with $7.6$;
both are near the {}``nominal'' value of the range in table \ref{tab:Number-of-LOC},
while the rest would appear upon or above the {}``complex'' end.
The remaining Java project, Eclipse, has the highest measure of all
but both versions of Emacs, with $14.1$. This value might be an artifact
of pruning (as explained in subsection \ref{sub:Pruning}); however,
all attempts at consolidation yielded the same results, or worse.
It appears that Eclipse is not as modular in high levels of componentization
as other packages; or that its logical structure is not reflected
in its on-disk layout.

\begin{figure}[th]
\begin{centering}
\includegraphics{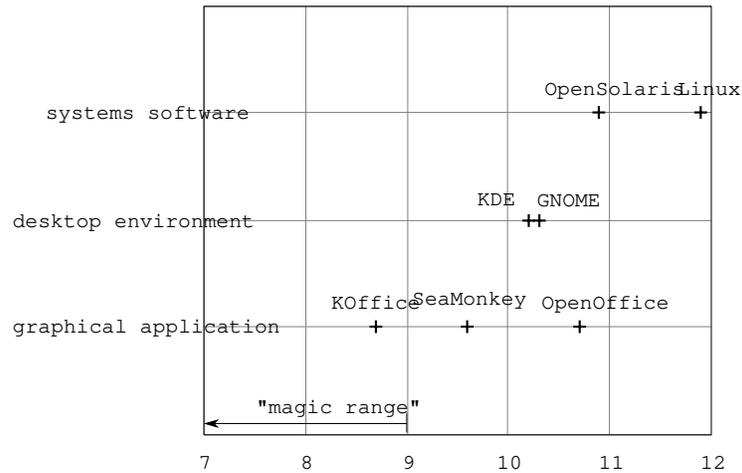}
\par\end{centering}

\caption{\label{fig:complexity-area}Complexity per application area. As should
be expected, systems software is more complex than graphical environments,
which in turn are more complex than graphical applications. The exception
is OpenOffice, a software package with roots in the eighties.}

\end{figure}

Emacs and XEmacs-Packages have far higher measures of complexity for
level > 1 than the rest. Even the more modern refactorization has
$18.8$, while the venerable branch has a staggering value of $48.4$.
Even if at low levels they behave much better, it appears that they
just are not modularized in a modern sense, i.e. in a hierarchical
fashion. As to the rest, they are just above the {}``magic range'',
from Linux with $11.9$ to KOffice with $8.7$. At least for the C/C++
packages there appears to be some correlation between application
area and complexity: systems software packages like Linux and OpenSolaris
(with $11.9$ and $10.9$ respectively) rank higher than the desktop
environments GNOME and KDE ($10.3$ and $10.2$), in turn higher than
the graphical applications SeaMonkey and KOffice ($9.6$ and $8.7$).
OpenOffice provides the exception for a graphical application with
$10.7$, but it is also the only package in the set with roots in
the decade of 1980 \cite{key-20}. Figure \ref{fig:complexity-area}
shows a summary of the effect. Incidentally, this correlation has
only been made apparent after pruning.

On the other end, level 0 complexity or $\mathsf{LOC}$ per function
obviously do not behave as subcomponents in a component: the value
is out of the {}``magic range'' for almost all software packages.
Java packages show the lowest count for $\mathsf{LOC}$ per function
with $8.6$, $8.6$and $9.5$; but the highest count is for OpenSolaris
with $35.6$, which otherwise shows good modularity. It appears that
having far more than $7\pm2\mathsf{LOC}$ per function is not a problem
for package modularity.

More surprising is to find that level 1 complexity (functions per
file) is out of range too in almost all instances, except for JBoss.
For C software results are $9.6$, $16.0$ and $16.4$, very similar
to C++ where we find $10.4$, $14.5$, $15.2$ and $15.7$. These
results are typical of the values that have been observed in other
software packages: a range of $[2,20]$ would capture most C and C++
software. Sometimes an effort is done to modularize code at this level,
as can be seen in OpenSolaris and OpenOffice. Java software seems
to be closer to this goal: results for methods per file are now $5.7$,
$11.2$ and $11.3$, mostly still out of range but close enough. In
wider field tests a range of $[2,12]$ would capture most Java software.
The remaining software in Lisp meanwhile does a little worse with
$17.6$ and $25.3$ functions per file; some efforts seem to have
been given in this front, but the amount of the reduction from Emacs
to XEmacs-Packages is not too significant (and is even less evident
in XEmacs proper, not included in this study).

A possible explanation for this discrepancy at level 1 is that only
exported functionality should be counted; when considering a component
from the outside a developer only needs to be concerned with visible
functions (i.e. public functions in C++, or public methods in Java).
The problem remains for internal consideration of components, which
is precisely where the {}``magic range'' can have the biggest effect.
Another explanation is that only in object-oriented languages there
is a sensibility for modularity within source files, and it appears
to be the case at least for Java. In the next subsection some explanations
are explored for C++.

It appears that many popular free software packages tend to be organized
as one would expect from purely theoretical grounds; after measuring
more than 25 million lines of code it has been found that several
popular free software packages tend to comply with the theoretical
framework about a hierarchy of components, at least as far as file
layout is concerned. Other software packages that were conceived using
less modern practices, like Emacs, do not follow this organization.
Whether this structure indeed corresponds to modularization is a different
question, difficult to answer without intimate knowledge of their
internal structure; more research is required to give definitive responses.

\subsection{Classes and Files in C++\label{sub:c++}}

C++ offers the possibility to arrange several classes in a file; according
to the theoretical study presented in subsection \ref{sub:Articulation},
this organization should result in another level of articulation within
files. It is however good coding advice to put only one module unit
per file \cite{key-8}; in C++ this translates into placing each class
in a separate file. If this convention is followed the number of functions
per file might be expected to follow the magic range. In Java, where
the limitation of declaring one class per file is enforced by the
language, the results are indeed near the expectation with $5.7$,
$11.2$ and $11.3$. However, C++ results ($10.4$, $14.5$, $15.2$
and $15.7$) are similar to those for C; again, a typical range seems
to be $[7,20]$. On the other hand, if each file contained two levels
of articulation (functions and classes) the number of functions per
file that could be expected would rather be $(7\pm2)^{2}=[25,81]$.
It is clear that in C++ there is nothing near this interval either.
This deviation deserves a deeper exploration. In this subsection four
possible explanations will be advanced.

The C++ convention is to have one \emph{public} class per file; there
can be some private classes within the same source file. One private
class per public class (or two classes per file) would yield an interval
twice the {}``magic range'', or $[10,18]$, consistent with observations.
Alternatively, the range might apply only to public functions and
not to those private to the class, but this explanation was discarded
in the previous subsection since external consideration is not where
the number of elements would have the most effect.

Some software packages may disregard the {}``one public class per
file'' convention altogether or in parts; sometimes it is possible
to find multi-class files in otherwise compliant software packages,
yielding more than one level of articulation between files and functions.
For example, an average of $1.3$ levels would yield an interval of
$(7\pm2)^{1.3}\approx[8,17]$, also congruent with observations. It
can be noted that mean depth for C packages is $4.85$, and $5.32$
for Java; but for C++ it is lower with an average of only $4.57$.
The difference may be suggestive that C++ may have a partial {}``hidden
articulation level'' in classes per file, and even grossly coincides
with the conjectured $1.3$ levels, but it is not statistically significant.
Note also that KOffice is quite smaller than most packages.

C++ also tends to require a large number of supporting code for each
class implemented. It is not uncommon to find an explicit constructor,
a destructor, an assignment operator and a copy constructor declared
for each class \cite{key-9}. With four additional functions per file,
the interval would become $[9,13]$; if half the functions were supporting
code, we would again get a range of $[10,18]$.

The last explanation is psychological, and can be the underlying cause
for all the previous hypotheses. It seems that C++ developers don't
see a problem with having as many as 20 functions per file; in fact
going below 10 may look very difficult when writing C++ code. For
example, Kalev writes \cite{key-10}:
\begin{quotation}
As a rule, a class that exceeds a 20-30 member function count is suspicious.
\end{quotation}
Such a number would raise the hairs on the back of the neck for experienced
Java developers. Whether this concern is valid or it is just an idiomatic
difference shall remain an open question until further research can
validate the alternatives.

\section{Conclusion}

The theoretical framework presented in this study should allow developers
to view really large systems in a new light; the experimental results
show that complexity in a hierarchy of components can be dealt with
and measured. Until now most of these dealings have been made intuitively;
hopefully the metrics developed in this paper can help developers
think about how to structure their software, and they can be used
for complexity analysis by third parties.

As a first proof of their usefulness, several object oriented software
packages have been shown to have a bigger sensibility to modularization
than those written in procedural languages, and these in turn higher
than functional language software. These concerns may explain their
relative popularity better than language-specific features. Application
area also seems to play a role, as does age of the project. Apparently
the set of metrics reflect, if not the exact hierarchy of logical
components, at least general modularity concerns.

\subsection{Metrics in the Real World}

Metrics can be highly enlightening, but if they cannot be computed
automatically they are less likely to be used in practice. Function
points are very interesting but hard to compute, and therefore they
have not yet reached their potential as a universal size measure \cite{key-6}.

The metrics presented here are very easy to compute; a few regular
expressions and some commands should be enough to get full results.
A software package which analyzes source code automatically and produces
a report should be very easy to develop; extending existing packages
should be even easier.

Despite the automatic nature of the metrics, a certain amount of manual
inspection of the intermediate results must always be performed in
order to assess if directory structure reflects software modularity.
This is usually the case with source code, so inspections are a good
idea in any case.

\subsection{Further Work}

This study has focused on well-known free software packages with sizes
within an order of magnitude ($500\mathsf{kLOC}$ to $5000\mathsf{kLOC}$).
Extending the study to a big range of projects (with hard data for
maintainability) would add reliability to the metrics proposed. Future
research will try to identify complex areas in otherwise modular programs.
An interesting line of research would be to model the software not
as an exponential tree, but as a more organic structure with outgrowths,
stubs and tufts; this approach might help locate problematic areas.

The set of metrics presented is only a first approximation to measures
of complexity. Until these metrics are validated by wider studies
they should not be applied to mission-critical projects or used to
make business decisions. On the other hand, making hierarchical complexity
measurable is an important step that should lead to more comprehensive
but hopefully just as simple metrics.

A goal that can immediately be applied to current development is to
make explicit and to document the hierarchical component organization.
The process of sorting out and naming levels of components following
a conscious and rational scheme will probably lead to better modularization;
in the end it is expected that more maintainable software will be
attained.

\subsection{Acknowledgments}

The author would like to thank Rodrigo Lumbreras for his insights
on professional C++ development, and Carlos Santisteban for his sharp
corrections.

\end{document}